\documentclass{webofc}
\usepackage[varg]{txfonts}   
\graphicspath{ {figs/} }
\usepackage{epsfig}

\allowdisplaybreaks

\begin{document}
\title{$(g-2)_\mu$ at four loops in QED}
%
%

\author{\firstname{Peter} \lastname{Marquard}\inst{1}\fnsep
  \and
  \firstname{Alexander V.} \lastname{Smirnov}\inst{2}\fnsep
  \and
  \firstname{Vladimir A.} \lastname{Smirnov}\inst{3}\fnsep
  \and
  \\
  \firstname{Matthias} \lastname{Steinhauser}\inst{4}\fnsep\thanks{\email{matthias.steinhauser@kit.edu}}
  \and
  \firstname{David} \lastname{Wellmann}\inst{4}\fnsep
}

\institute{
  Deutsches Elektronen-Synchrotron, DESY,
  15738 Zeuthen, Germany
  \and
  Research Computing Center, Moscow State University,
  119991, Moscow, Russia
  \and
  Skobeltsyn Institute of Nuclear Physics of Moscow State University,
  119991, Moscow, Russia
  \and
  Institut f{\"u}r Theoretische Teilchenphysik, Karlsruhe
  Institute of Technology (KIT),\\ 76128 Karlsruhe, Germany
}

\abstract{%
  We review the four-loop QED corrections to the 
  anomalous magnetic moment of the muon.
  The fermionic contributions with closed electron and
  tau contributions are discussed. Furthermore, 
  we report on a new independent calculation of
  the universal four-loop contribution and
  compare with existing results.
}
\maketitle
%


\section{Introduction}

The anomalous magnetic moment of the muon,
which is usually written as $a_\mu=(g-2)_\mu/2$,
measures the deviation from Dirac's prediction $g=2$.
Experimentally it is known with high precision
from measurements at BNL~\cite{Bennett:2006fi,Roberts:2010cj}
\begin{eqnarray}
  a_{\mu}^{\rm exp} &=& 116\,592\,089(63) \times 10^{-11}
  \,.
  \label{eq::amuexp}
\end{eqnarray}
It is expected that the uncertainty will be reduced in the coming
years. Actually, there are two experiments which are currently under
construction, one at Fermilab and one at J-PARC~\cite{Carey:2009zzb,Roberts_2017,Mibe_2017}

Also on the theory side an impressive precision has been reached. However,
since many years there is a persistent discrepancy of the order of
about 3~sigma.
The uncertainty of the theory prediction is dominated by the hadronic
contributions, both from the vacuum polarization~\cite{Teubner_2017,Zhang_2017,Jegerlehner_2017} (see
Refs.~\cite{Hagiwara:2011af,Jegerlehner:2017lbd,Davier:2017zfy} for a recent compilations)
and the so-called light-by-light part~\cite{Nyffeler_2017}.

The numerically largest contribution to $a_\mu$ is given by the QED part which
is known up to five loops. One-, two- and three-loop corrections are known
analytically from
Refs.~\cite{Schwinger:1948iu,Petermann:1957hs,Sommerfeld:1958,Laporta:1996mq}
and four- and five-loop contributions have been computed in
Refs.~\cite{Kinoshita:2004wi,Aoyama:2007mn,Aoyama:2012wk,Aoyama:2014sxa} using
numerical methods. The fermionic contributions involving closed tau and
electron loops have been cross checked in
Refs.~\cite{Kurz:2013exa,Kurz:2015bia,Kurz:2016bau}.  Very recently,
semi-analytic results for the universal contribution, i.e., the purely
photonic and muon-loop contribution, have been obtained in a remarkable
calculation by Laporta~\cite{Laporta:2017okg}.  It is based on an evaluation
of Feynman integrals with high-precision (several thousand digits) which was
enough to reconstruct rational coefficients of known transcendental constants
with the help of the PSLQ algorithm~\cite{PSLQ}.  In addition, there were
several contributions which were not recognized as known constants.  The final
result for the four-loop contribution to $a_\mu$ from~\cite{Laporta:2017okg}
is known to 1100 digits.

In this work we present results of an independent calculation
of the universal contribution.

In order to fix the notation we provide numerical
results for $a_\mu$ up to five-loop order which are given by
(numbers are taken from Refs.~\cite{Aoyama:2012wk,Aoyama:2014sxa})
\begin{eqnarray}
  a_\mu &=& \frac{(g-2)_\mu}{2} 
  \nonumber\\
  &=&
  \frac{\alpha}{2\pi}
  \label{eq::amu}\\
  &&
  + (-0.328\,478\ldots + 1.094\,336\ldots|_{e,\tau})   \left(\frac{\alpha}{\pi}\right)^2
  + (1.181\,241\ldots + 22.869\,268\ldots|_{e,\tau})   \left(\frac{\alpha}{\pi}\right)^3
  \nonumber\\
  &&
  + (-1.912\,98(84) + 132.790\,3(60)|_{e,\tau}) \left(\frac{\alpha}{\pi}\right)^4
  + (9.168(571) + 744.123(870)|_{e,\tau})      \left(\frac{\alpha}{\pi}\right)^5
  \,,
  \nonumber
\end{eqnarray}
where the ellipses indicate that the numbers are truncated and
actually more digits are known.
The universal part (first number in the brackets) has been
separated from electron and tau contributions (second number), which
appears for the first time at two loops. Note that the latter is
numerically dominant due to unsuppressed large
logarithms of the ratio of the electron and muon mass,
$\log(m_\mu/m_e)\approx 5.332$. At $\ell$-loop order such logarithms
occur up to the $(\ell-1)$th order. On the other hand, heavy virtual
particles are decoupled and thus the tau contributions are suppressed
by $m_\mu^2/m_\tau^2$. They are numerically small.

Let us note that up to terms suppressed by $m_e^2/m_\mu^2$
the first numbers in the coefficients of Eq.~(\ref{eq::amu})
coincide with the anomalous magnetic moment of the electron, $a_e$.

It is interesting to note that after inserting the fine structure
constant the four-loop coefficient evaluates to
\begin{eqnarray}
  a_\mu^{(8)} &=& (-1.912\,98 + 132.790\,3|_{e,\tau})
  \left(\frac{\alpha}{\pi}\right)^4 
  \,\,\approx\,\,
  381 \times 10^{-11}
  \,,
\end{eqnarray}
which is of the same order of magnitude as the current difference
between the Standard Model prediction of $a_\mu$ and the experimental
value given in Eq.~(\ref{eq::amuexp}). Furthermore, it is larger than
the uncertainties of the hadronic vacuum polarization and
light-by-light contributions which are both of the order of $40 \times
10^{-11}$.  Thus, an independent cross check of the four-loop QCD
contributions is indispensable.


\section{Technical remarks}

The techniques used to obtain the results in
Refs.~\cite{Kurz:2013exa,Kurz:2015bia,Kurz:2016bau} and for the universal
part, which we report below, have largely been developed in the context of the
$\overline{\rm MS}$-on-shell quark mass relation in QCD. To obtain the mass
relation one has to evaluate on-shell integrals up to four loops which are
also the basis for the anomalous magnetic moment. In fact, we use the same
integral families as defined in Refs.~\cite{Marquard:2015qpa,Marquard:2016dcn}
and express the four-loop expression for $a_\mu$ as a linear combination of
scalar integrals.  The latter are reduced to master integrals with the help of
{\tt FIRE}~\cite{Smirnov:2014hma} and {\tt                                  
  Crusher}~\cite{crusher}. Let us mention that the reduction of the
integrals contributing to $a_\mu$ is more expensive since vertex integrals
(instead of two-point functions) are considered which are expanded around
vanishing momentum transfer of the photon.  Thus, in the corresponding
integrals the total power of the propagators is increased by at least two as compared
to the integrals needed for the $\overline{\rm MS}$-on-shell relation.  For
the $\overline{\rm MS}$-on-shell relation we have to evaluate 386 master
integrals; a subset of 357 master integrals contribute to $a_\mu$. For
details concerning their evaluation we refer to
Ref.~\cite{Marquard:2016dcn}. To obtain the precision mentioned below some of
the master integrals had to be evaluated with higher precision following the
methods of~\cite{Marquard:2016dcn}.

Additional work is needed for the fermionic contributions with closed
electron or tau loops. In both cases it is appealing to perform an
asymptotic expansion either for $m_e\ll m_\mu$ or $m_\mu \ll
m_\tau$. The latter is a Euclidean-like asymptotic expansion which can
be performed with the help of the program {\tt
  exp}~\cite{Harlander:1997zb,Seidensticker:1999bb}. The most
complicated integrals which have to be evaluated are four-loop vacuum
integrals which are well studied in the literature (see, for example,
Ref.~\cite{Marquard:2016dcn} and references therein). All other contributions
are of lower loop order and also known analytically. Thus, the
four-loop contribution to $a_\mu$ containing tau leptons is known
analytically as a series in $m_\mu/m_\tau$ which is rapidly
converging~\cite{Kurz:2013exa}.

To obtain an expansion of the electron-loop contribution in
$m_e/m_\mu$ an asymptotic expansion around the on-shell limit has to be
performed. The complicated integrals one has to compute are either of
on-shell type (as for the universal contribution) or integrals which
contain linear propagators of the form $1/p\cdot q$ where
$q^2=m_\mu^2$ is the external momentum and $p$ is a loop
momentum. Some integrals of this type can be computed analytically,
others are computed numerically using {\tt
  FIESTA}~\cite{Smirnov:2013eza}. In
Refs.~\cite{Kurz:2015bia,Kurz:2016bau} expansion terms up to order
$m_e^3/m_\mu^3$ have been computed which show a good convergence
behaviour. Let us remark that the numerically dominant contribution
arises from the light-by-light-type contributions which have been
computed in~\cite{Kurz:2015bia}.

In Refs.~\cite{Kinoshita:2004wi,Aoyama:2007mn,Aoyama:2012wk}
a completely different technique has been used
to compute the four-loop corrections to $a_\mu$. In a first step a
finite expression is constructed by generating the proper counterterms
together with four-loop diagrams which is afterwards
integrated numerically.

In Ref.~\cite{Laporta:2017okg}, similar to our approach, all occurring
integrals are reduced to a small set of master integrals. However,
different software is used and most probably also a different basis of
master integrals is chosen. Furthermore,
Ref.~\cite{Laporta:2017okg} manages to obtain high-precision numerical
expressions for all master integrals whereas we have chosen a more
automated approach and stopped manipulating the integrals once the
desired precision has been reached.


\section{Results}

Let us start with discussing the universal part to $a_\mu$ which
consists of the pure photon contribution and the contribution with
closed muon loops. It can be subdivided into six gauge invariant
subsets; a representative diagram for each one is shown in the first
column of Tab.~\ref{tab::amu}. The second column in
Tab.~\ref{tab::amu} contains the corresponding results from
Ref.~\cite{Aoyama:2012wj}, this work, and Ref.~\cite{Laporta:2017okg},
respectively (from top to bottom).
The results from Ref.~\cite{Aoyama:2012wj} are taken from Table~I
of that reference and the uncertainties are added in quadrature
in case several contributions had to be combined. The uncertainty of
the results obtained in this work are the quadratically combined
results from the individual $\epsilon$ coefficients of the master
integrals. We refrain from introducing a ``security factor''
(as, e.g., in Ref.~\cite{Marquard:2016dcn}) for the universal contribution 
since the four-loop result for
$a_\mu$ has also been computed by two other groups.
There is no uncertainty in the result provided in
Ref.~\cite{Laporta:2017okg}. 


\begin{table}[t]
  \centering
  \begin{tabular}{cc}
    \hline
    Representative
    &
    Contribution of $a_\mu$
    \\
    Feynman diagram
    &
    \\ \hline
    \raisebox{-2em}{\includegraphics[width=4em]{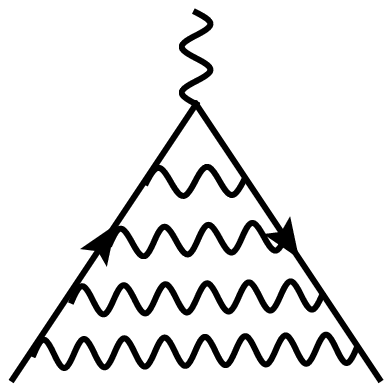}}
    & 
    \begin{minipage}{25em}
      $-2.1755 \pm 0.0020$\\
      $-2.161 \pm 0.065$\\
      $- 2.176866027739540077443259355895893938670$
    \end{minipage}
    \\ \hline
    \raisebox{-2em}{\includegraphics[width=4em]{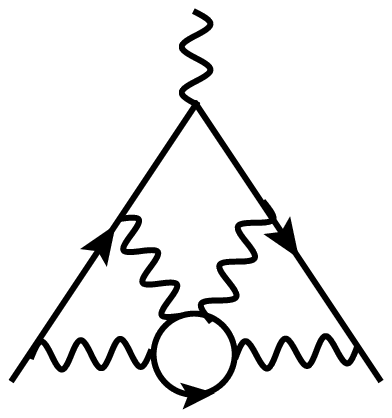}}
    & 
    \begin{minipage}{25em}
      $0.05596\pm 0.0001$\\
      $0.077\pm 0.031$\\
      $0.056110899897828364831469274418908842233$
    \end{minipage}
    \\ \hline
    \raisebox{-2em}{\includegraphics[width=4em]{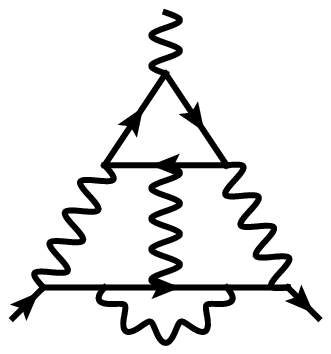}}
    & 
    \begin{minipage}{25em}
      $-0.3162\pm 0.0002$\\
      $-0.3048 \pm 0.021$\\
      $-0.316538390648940158843260382381513284828$
    \end{minipage}
    \\ \hline
    \raisebox{-2em}{\includegraphics[width=4em]{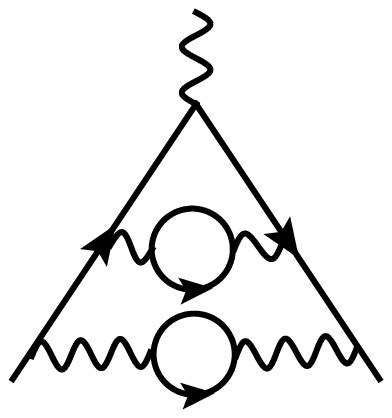}}
    & 
    \begin{minipage}{25em}
      $-0.074665 \pm 0.000006$\\
      $-0.07461  \pm 0.00008$\\
      $-0.074671184326105513860159965722793126809$
    \end{minipage}
    \\ \hline
    \raisebox{-2em}{\includegraphics[width=4em]{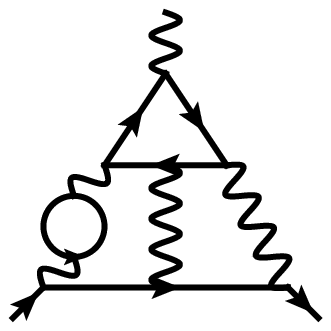}}
    & 
    \begin{minipage}{25em}
      $0.598838 \pm 0.000019$\\
      $0.597204\pm 0.0012$\\
      $0.598842072031421820464649513201747727836$
    \end{minipage}
    \\ \hline
    \raisebox{-2em}{\includegraphics[width=4em]{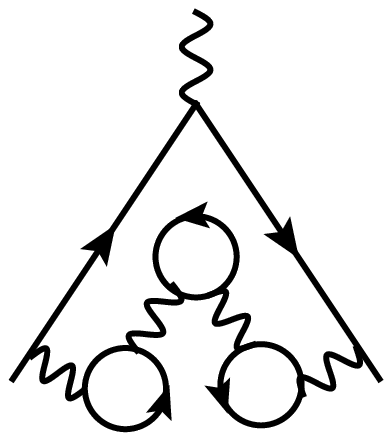}}
    & 
    \begin{minipage}{25em}
      $0.000876865858889990697913748939713726165$\\
      $0.000876865858889990697913748939713726165$\\
      $0.000876865858889990697913748939713726165$
    \end{minipage}
    \\ \hline
 \end{tabular}
  \caption{
    \label{tab::amu}
    The three numbers given in each row
    (from top to bottom) are taken from 
    \protect\cite{Aoyama:2012wj}, 
    this work, and 
    \protect\cite{Laporta:2017okg},
    respectively.}
\end{table}


Within the given uncertainties the results from~\cite{Aoyama:2012wj}
and this work agree with the semi-analytic expressions
of~\cite{Laporta:2017okg}. In most cases our uncertainty is at the per
cent level or below, except for the contribution in the second row
where a 40\% uncertainty is observed. Note, that the absolute size of
the uncertainty is of the same order as the one in the first and third
row. However, due to cancellations from individual contributions, the
central value is significantly smaller.

In the following we summarize the four-loop QED
contributions and compare the results from
the different groups. Denoting the coefficient of
$(\alpha/\pi)^4$ by $a_\mu^{(8)}$ we have
\begin{align}
  & \mbox{} \hspace*{3em} \mbox{universal} \hspace*{3em} \mbox{$e^-$} \hspace*{7em}
  \mbox{$\tau$} \hspace*{5em} \mbox{$e^- + \tau$} 
  \nonumber\\
  &a_\mu^{(8)} = -1.87(12)
  \hphantom{iiiiii}
  + 132.86(48) 
  \hphantom{xx}
  + 0.0424941(53) + 0.062722(10) 
  &&\mbox{this work and~\cite{Kurz:2013exa,Kurz:2015bia,Kurz:2016bau}}
  \nonumber\\
  &a_\mu^{(8)} = -1.912\,98(84)
  + 132.6852(60) + 0.04234(12) 
  \hphantom{xx}
  + 0.06272(4)
  &&\mbox{\cite{Aoyama:2012wk}}
  \nonumber\\
  &a_\mu^{(8)} = -1.9122457649264\ldots
  &&\mbox{\cite{Laporta:2017okg}}
  \nonumber
\end{align}
Note that the uncertainties in the first line in the parts
involving a tau lepton are due to the lepton masses only.
After multiplication with $(\alpha/\pi)^4$ we obtain
for the three equations
\begin{align}
  & (-5.44(35) 
  \hphantom{iiiiiii}
  + 386.77(1.40) + 0.12371(15) + 0.182592(29))
  \times  10^{-11}
  &&
  \mbox{this work and~\cite{Kurz:2013exa,Kurz:2015bia,Kurz:2016bau}}
  \nonumber\\
  & (-5.56894(245) + 386.264(17) 
  \hphantom{i}
  + 0.12326(35) + 0.18259(12))
  \times  10^{-11}
  &&
  \mbox{\cite{Aoyama:2012wk}}
  \nonumber\\
  & (-5.56679893738506\ldots + \ldots )\times  10^{-11}
  &&
  \mbox{\cite{Laporta:2017okg}}
  \nonumber
\end{align}
The uncertainty of our result is about two orders of magnitudes
larger. It is nevertheless much smaller than the current and foreseen
uncertainties from both experiment and the hadronic contributions.
This can be seen by considering
the difference between the
experimental result and the Standard Model prediction
which is given by (see, e.g., Ref.~\cite{Aoyama:2012wk})
\begin{eqnarray*}
  a_\mu({\rm exp}) - a_\mu({\rm SM}) &\approx& 250(90) \times 10^{-11}
  \,.
\end{eqnarray*}
The uncertainty is about two orders of magnitude larger than
our numerical uncertainty cited above. This remains even true
after applying the improvements by a factor~4.
Thus, it can be claimed that the
four-loop contribution for $a_\mu$ is cross-checked:
There are three independent calculations for the universal part
and the electron and tau contributions have been
computed by two independent groups.

Let us finally remark on $a_e$.
The Standard Model prediction given in Ref.~\cite{Laporta:2017okg}
reads
\begin{eqnarray}
  a_e({\rm SM}) &=& {115\,965\,218.1664(23)(16)(763) \times
    10^{-11}}
  \,,
\end{eqnarray}
where the three uncertainties have their origin in the
numerical accuracy of the five-loop calculation, the hadronic and
electroweak corrections and the fine structure constant.
Due to the result of Ref.~\cite{Laporta:2017okg}
an additional uncertainty of ``(60)'', which is still present
in~\cite{Aoyama:2012wj}, has been removed.
Note that our result for the universal part of $a_\mu$
can also be applied to $a_e$. However, since it has an uncertainty which is
two orders of magnitude larger than the one cited in~\cite{Aoyama:2012wj}
it is not competitive to~\cite{Aoyama:2012wj} and~\cite{Laporta:2017okg}.


\section{Conclusions}

We summarize all four-loop QED contributions to the anomalous magnetic
moment of the muon.  They have been computed for the first time in
Refs.~\cite{Kinoshita:2004wi,Aoyama:2007mn,Aoyama:2012wk}. An
independent cross check of the tau-loop contributions can be found in
Ref.~\cite{Kurz:2013exa} where analytic results are provided for the
expansion in $m_\mu/m_\tau$. The electron-loop contributions have been
cross checked in Refs.~\cite{Kurz:2015bia,Kurz:2016bau} where an
asymptotic expansion in $m_e/m_\mu$ has been used. An independent
semi-analytic calculation of the universal (purely photonic and
muon-loop) contribution has been obtained in
Ref.~\cite{Laporta:2017okg}. In this work we provide yet another
independent cross check.  In summary, all four-loop QED contributions to
$a_\mu$ have been computed by at least two groups independently using
completely different methods.


\section*{Acknowledgments}
We thank the High Performance Computing Center Stuttgart (HLRS) and
the Supercomputing Center of Lomonosov Moscow State University for
providing computing time used for the numerical computations with {\tt
  FIESTA}.
P.M. was supported in part by the EU Network HIGGSTOOLS PITN-GA-2012-316704.

\vspace*{-1em}



\begin{thebibliography}{00}  

%
%

\bibitem{Bennett:2006fi}
  G.~W.~Bennett {\it et al.}  [Muon G-2 Collaboration],
  Phys.\ Rev.\ D {\bf 73} (2006) 072003
  [hep-ex/0602035].

\bibitem{Roberts:2010cj}
  B.~L.~Roberts,
  Chin.\ Phys.\ C {\bf 34} (2010) 741
  [arXiv:1001.2898 [hep-ex]].

\bibitem{Carey:2009zzb}
  R.~M.~Carey {\it et al.},
  FERMILAB-PROPOSAL-0989.

\bibitem{Roberts_2017}
B. Lee Roberts, these proceedings.

\bibitem{Mibe_2017}
T. Mibe, these proceedings.

\bibitem{Teubner_2017}
T. Teubner, these proceedings.

\bibitem{Zhang_2017}
Z. Zhang, these proceedings.

\bibitem{Jegerlehner_2017}
F. Jegerlehner, these proceedings.

\bibitem{Hagiwara:2011af}
  K.~Hagiwara, R.~Liao, A.~D.~Martin, D.~Nomura and T.~Teubner,
  J.\ Phys.\ G {\bf 38} (2011) 085003
  doi:10.1088/0954-3899/38/8/085003
  [arXiv:1105.3149 [hep-ph]].

\bibitem{Jegerlehner:2017lbd}
  F.~Jegerlehner,
  arXiv:1705.00263 [hep-ph].

\bibitem{Davier:2017zfy}
  M.~Davier, A.~Hoecker, B.~Malaescu and Z.~Zhang,
  arXiv:1706.09436 [hep-ph].

\bibitem{Nyffeler_2017}
A. Nyffeler, these proceedings.

\bibitem{Schwinger:1948iu}
  J.~S.~Schwinger,
  Phys.\ Rev.\  {\bf 73} (1948) 416.
  doi:10.1103/PhysRev.73.416

\bibitem{Petermann:1957hs}
  A.~Petermann,
  Helv.\ Phys.\ Acta {\bf 30} (1957) 407.

\bibitem{Sommerfeld:1958}
C. M. Sommerfield, Ann. Phys. (N.Y.) 5 (1958) 26.

\bibitem{Laporta:1996mq}
  S.~Laporta and E.~Remiddi,
  Phys.\ Lett.\ B {\bf 379} (1996) 283
  [hep-ph/9602417].

\bibitem{Kinoshita:2004wi}
  T.~Kinoshita and M.~Nio,
  Phys.\ Rev.\ D {\bf 70} (2004) 113001
  [hep-ph/0402206].

\bibitem{Aoyama:2007mn}
  T.~Aoyama, M.~Hayakawa, T.~Kinoshita and M.~Nio,
  Phys.\ Rev.\ D {\bf 77} (2008) 053012
  [arXiv:0712.2607 [hep-ph]].

\bibitem{Aoyama:2012wk}
  T.~Aoyama, M.~Hayakawa, T.~Kinoshita and M.~Nio,
  Phys.\ Rev.\ Lett.\  {\bf 109} (2012) 111808
  [arXiv:1205.5370 [hep-ph]].

\bibitem{Aoyama:2014sxa}
  T.~Aoyama, M.~Hayakawa, T.~Kinoshita and M.~Nio,
  Phys.\ Rev.\ D {\bf 91} (2015) no.3,  033006
   Erratum: [Phys.\ Rev.\ D {\bf 96} (2017) no.1,  019901]
  doi:10.1103/PhysRevD.91.033006, 10.1103/PhysRevD.96.019901
  [arXiv:1412.8284 [hep-ph]].

\bibitem{Kurz:2013exa}
  A.~Kurz, T.~Liu, P.~Marquard and M.~Steinhauser,
  Nucl.\ Phys.\ B {\bf 879} (2014) 1
  [arXiv:1311.2471 [hep-ph]].

\bibitem{Kurz:2015bia}
  A.~Kurz, T.~Liu, P.~Marquard, A.~V.~Smirnov, V.~A.~Smirnov and
  M.~Steinhauser,
  Phys.\ Rev.\ D {\bf 92} (2015) 7,  073019
  doi:10.1103/PhysRevD.92.073019
  [arXiv:1508.00901 [hep-ph]].

\bibitem{Kurz:2016bau}
  A.~Kurz, T.~Liu, P.~Marquard, A.~Smirnov, V.~Smirnov and M.~Steinhauser,
  Phys.\ Rev.\ D {\bf 93} (2016) no.5,  053017
  doi:10.1103/PhysRevD.93.053017
  [arXiv:1602.02785 [hep-ph]].

\bibitem{Laporta:2017okg}
  S.~Laporta,
  Phys.\ Lett.\ B {\bf 772} (2017) 232
  doi:10.1016/j.physletb.2017.06.056
  [arXiv:1704.06996 [hep-ph]].

\bibitem{PSLQ}
H.R.P.~Ferguson and D.H.~Bailey, RNR Technical Report, RNR-91-032;
H.R.P.~Ferguson, D.H.~Bailey and S.~Arno, NASA Technical Report,
NAS-96-005.

\bibitem{Marquard:2015qpa}
  P.~Marquard, A.~V.~Smirnov, V.~A.~Smirnov and M.~Steinhauser,
  Phys.\ Rev.\ Lett.\  {\bf 114} (2015) 14,  142002
  [arXiv:1502.01030 [hep-ph]].

\bibitem{Marquard:2016dcn}
  P.~Marquard, A.~V.~Smirnov, V.~A.~Smirnov, M.~Steinhauser and D.~Wellmann,
  Phys.\ Rev.\ D {\bf 94} (2016) no.7,  074025
  doi:10.1103/PhysRevD.94.074025
  [arXiv:1606.06754 [hep-ph]].

\bibitem{Smirnov:2014hma}
  A.~V.~Smirnov,
  Comput.\ Phys.\ Commun.\  {\bf 189} (2014) 182
  [arXiv:1408.2372 [hep-ph]].

\bibitem{crusher}
P.~Marquard, D.~Seidel, unpublished.

\bibitem{Harlander:1997zb}
  R.~Harlander, T.~Seidensticker and M.~Steinhauser,
  Phys.\ Lett.\ B {\bf 426} (1998) 125,
  arXiv:hep-ph/9712228.

\bibitem{Seidensticker:1999bb}
  T.~Seidensticker,
  arXiv:hep-ph/9905298.

\bibitem{Smirnov:2013eza}
  A.~V.~Smirnov,
  Comput.\ Phys.\ Commun.\  {\bf 185} (2014) 2090
  [arXiv:1312.3186 [hep-ph]].

\bibitem{Aoyama:2012wj}
  T.~Aoyama, M.~Hayakawa, T.~Kinoshita and M.~Nio,
  Phys.\ Rev.\ Lett.\  {\bf 109} (2012) 111807
  [arXiv:1205.5368 [hep-ph]].


\end{thebibliography}
\end{document}